\documentstyle[12pt]{article}
\begin{document}
\begin{center}
{\Large \bf
Phase structure and confinement properties of noncompact
gauge theories I }  \\
\vspace{1cm}
{\large O.A.~Borisenko, V.K.~Petrov, G.M.~Zinovjev
\footnote{email: gezin@gluk.apc.org}}\\
{\large \it
N.N.Bogolyubov Institute for Theoretical Physics, National Academy
of Sciences of Ukraine, 252143 Kiev, Ukraine}\\
\vspace{1cm}
{\large J.~Boh\'a\v cik,
\footnote{email: bohacik@savba.savba.sk}}\\
{\large \it
Institute of Physics, Slovak Academy of Sciences,
842 28 Bratislava, Slovakia}\\
\end{center}
\vspace{.5cm}

\begin{abstract}
In the context of reviewing noncompact lattice gauge models at zero
and finite temperature we study in detail a contribution of
the invariant measure and the time-like plaquette configurations
to correlation functions, analyze the problem of the compactness
of the potentials in respect to the confinement and indicate
the essential features to deal with the Wilson gauge theory in the weak
coupling region. A method for calculating an effective confining
noncompact model is also proposed.
\end{abstract}

\newpage

\section{Introduction}

Since K.~Wilson proposed \cite{wilson} twenty years ago to quantize the field
theory on a lattice in the Euclidean space-time with an exact gauge
invariance in order to make the strong coupling calculations, the lattice
approach combined especially with the numerical Monte Carlo simulations
has provided a huge progress of the quantum chromodynamics (QCD). However,
the problem of a confinement mechanism did not become less intriguing
because a conceptually simple mechanism in strong coupling regime
could not proceed along the same line to the continuum theory. The
principal difficulties here is to define the proper configurations
(monopoles, vortices,...) of the compact lattice gauge fields
which are the most essential ones for forming the confining forces,
and how one may identify them in the continuum theory?

In this paper we make an attempt to advance this question again constructing
an effective noncompact model starting from the lattice gauge theory (LGT)
and aiming to analyze the weak coupling region. Generally speaking one
should presuppose the existence of the only mechanism of confinement
both on the lattice and in the continuum, in order to address this problem
unambiguously.
Because it is not obvious, one must argue the proffered statement.
We shall come back to this point later, accepting this as conjecture for now.
Besides, we need to develop an ingenious approach to construct an adequate
quantum theory of noncompact potentials starting from compact LGT.

Seems, such a possibility does exist. Exploring the lattice models which
are analytically solvable in a sense and exhibit the confinement property,
we could identify the important configurations and construct their
correct noncompact limit (of course, if it exists). This is the
principal strategy of the following paper. As the first step, we elaborate
the chromoelectric part of the Wilson action (WA) which states in this
approach the existence of confining forces in the low temperature phase.
However, the quantum noncompact lattice theory based on the naive
limit \footnote{Throughout this paper we use the following
limits of the lattice theory: 1) naive limit
means the expansion of lattice gauge field matrices around the unit
matrix and leads to the classical (lattice or continuum)
Yang-Mills action; 2) perturbative limit of Wilson LGT is taken
as a limit of vanishing lattice spacing together with
small coupling constant expansion and conventional
renormalization procedure done. This limit coincides with the
corresponding perturbative expansion of the continuum theory; 3)
nonperturbative limit is defined as a limit $a \rightarrow 0$ together
with a proper renormalization procedure after integrating the
lattice partition function over configurations of the gauge fields
which are far from the unit matrix and hence are missed in the perturbative
expansion;
4) mentioning a noncompact (perturbative or nonperturbative) limit
we mean an effective action in terms of noncompact gauge
potentials obtained in the limit of a small coupling constant after a
partial summation over compact gauge fields.} of the chromoelectric part
of WA does not possess this property compelling to analyze a
nonperturbative limit of the model.  (One may worry at this point that
dealing with the chromoelectric part of the WA we are trapped by the
strong coupling region which could be far away from the continuum
limit. At this stage, however, our goal is to find a
noncompact {\it lattice} theory which would belong to the
same universality class as the compact LGT. Continuum limit
can be accomplished after including the chromomagnetic
part of WA in our scheme).

The strong confirmation that in such a way we may achieve a desirable
effective noncompact model (and even to find its continuum quantum limit)
one finds in \cite{pol,polrev}, where it was shown that
confinement can be obtained in the effective model
for $A_{0}$ gauge field only and, in the first approximation,
we may set space-gauge fields $A_{n}$ equal to zero.
In fact, the effective action given in Refs. \cite{pol,polrev} shares the
same important features as an effective theory for the time-component
of gauge field $A_0$, which is calculated from the chromoelectric lattice
action.

Another problem closely connected with forementioned is the
deconfinement phase transition, which takes place in compact lattice theory
at finite temperature. As is known, the chromoelectric part of the compact
action is well indicative again, exhibiting the deconfinement.

Advertising the worthwhile results of our investigations,
we would like to mention a construction of a nonperturbative noncompact
limit of the Wison model in the weak coupling region and evaluation of the
corresponding effective model. What we obtained differs from the naive
noncompact generalization of the Yang-Mills theory because it includes
$Z(N)$ symmetry of WA and an influence of the invariant measure (as
specified below). We demonstrate that the mechanism of confinement in
the model developed is essentially the same, as that in the initial compact
theory and the new ingredients of noncompact formulation are playing
the crucial role to have confinement available. A string tension is
evaluated in the model and a generalization to include the chromomagnetic
part of WA is argued. Certainly, it does not solve the confinement
problem but permits to have a noncompact formulation on the same
footing as compact LGT.

We are going to present these results in two articles. The present
paper is organized as follows. In sect. 2 we remind briefly the
decisive features of the compact LGT in the strong coupling limit.
Sect. 3 is devoted to discussion of the noncompact lattice models and their
connection, both with the compact ones and with the continuum Yang-Mills
theory. We construct
and analyze noncompact model with compact $A_0$ integration in sect. 4.
In sect. 5 we discuss an effective way to include an invariant group
measure into noncompact models. We close in sect. 6 with a discussion
of compactness problem of potentials and give simple examples how
compactification could lead to the linear potential between probe
charges. On the other hand we claim it is quite enough in some models
with compact variables to perform noncompact Gaussian integration only
over dominating configurations to achieve linear potential. The main
issue of this discussion is a demonstration of how noncompact theory
can confine in the same way that compact theory does.

\section{The essentials of strong coupling compact LGT}

Let us consider compact $SU(N)$ and $SU(N)/Z(N)$
gauge theories on the lattice. The Wilson formulation of LGT
has the following form \cite{wilson}
\begin{equation}
Z=\int D\mu (U)  \exp (\lambda \sum_{p} \Omega(\partial p)),
\label{lgtc}
\end{equation}
\noindent
where $\Omega$ is a character of the fundamental representation
of a compact Lie group $G$, $D \mu (U)$ is the invariant
integration measure and $\lambda =\frac{2N_{c}}{g^{2}}$.
We would like to recall now some properties
of the Wilson LGT which will be essential here. The majority of
exact analytical results obtained by studying  the theory
(\ref{lgtc}) were achieved by strong coupling expansion.
These results in respect to confining properties are usually
related to the Wilson criterion of confinement, expressed by
the area law for the Wilson loops, which are non-trivial on the $Z(N)$
subgroup \cite{wilson}:
If the Wilson loop $\Omega_{\nu}(\partial C) = Tr \prod_{l \in C}
U_{\nu}(l)$ obeys the area behaviour
\begin{equation}
<\Omega_{\nu}(\partial C)>  \leq K_{0} \exp (-K_{1} area(C)),
\label{area}
\end{equation}
\noindent
the static colour charges in representation containing $Z(N)$
will be confined. It was proved in \cite{seil}, that
$<\Omega_{\nu}(\partial C)>$ shows the area behaviour in
a region of a convergence of the strong coupling expansion
for the $SU(N)$ gauge group. At the same time, the Wilson loop in the
adjoint representation obeys the perimeter law \cite{seil}.
What are the mechanisms of the such behaviour?
Two possibilities mainly dominate through
the discussions - monopole condensation and vortex condensation.

Here we are sticking to the opinion that the status of the vortex
condensation mechanism is better analytically founded, at least in
the strong coupling region thus preferring the special $Z(N)$ configurations
contributing to path integral to provide a confinement. In \cite{mack}
a sufficient condition for confinement by $Z(N)$ vortex condensation was
derived. $Z(N)$ vortices there take a special form of $Z(N)$
singular transformations performed over a two-dimensional closed  surface,
and their condensation means that they must become ``fat'' in a certain way.
A direct calculation up to very high orders in $\lambda \sim g^{-2}$
confirms the expected behaviour of the condensate in pure $SU(2)$
gluodynamics according to the mentioned theorem \cite{mack,munst}.
It was also proven \cite{munst}, that a coefficient at the area law
for the vortex free energy exactly equals the string tension.
Indeed, the string tension (coefficient $K_1$ in (\ref{area})) calculated
from the vortex condensate is in accordance with MC data in the region
of the strong and the intermediate coupling. The evaluation of the vortex
condensate in the weak coupling region can be found in \cite{mw}.
A crucial feature of this mechanism is the breakdown of the
$SU(N)$ local gauge symmetry up to its $Z(N)$ local subgroup
\cite{mack2}. This dynamical Higgs mechanism leads to the
long-interacting forces between colour charges, disordered
behaviour of the Wilson loop and screening of all the gluonic states.

Believing in this strong coupling confinement picture and taking into
account that there is no phase transition at zero temperature
in $SU(2)$ and $SU(3)$ gauge theories, one may hope that this mechanism
would persist in the continuum theory as well, if $Z(N)$  configurations
survive the transition to the weak coupling regime. Certainly, it is not
the case at the naive continuum limit and at the continuum
limit of perturbative expansion. Therefore, the nonperturbative limit must
be studied. It has been demonstrated in \cite{stul}
that a nonperturbative continuum limit of the $SU(N)$ LGT
contains $Z(N)$-vortices already in a bare Lagrangian.
It is interesting to note that the monopole configurations
do not contribute to such not naive continuum limit \cite{stul}.
These facts are heuristically important though the theory having been
exposed in \cite{stul} is different from the conventional Yang-Mills one
(see sect. 6 for more discussion of this point).
Usual objection against this
$Z(N)$ confinement mechanism comes from the observation that
the Wilson loop in the adjoint representation shows presumably the area law
behaviour in the limit $N_{c} \rightarrow \infty$ ($N_{c}$ is a number of
colours) despite this representation does not feel $Z(N)$ variables.
This objection has been discussed in \cite{stul,stul1}
and we refer the interested readers to that discussion.
It is worth mentioning that, in fact, the monopole mechanism of confinement
runs into similar problems as well \cite{greensite}; we do not know
whether there exist any answers to the questions put forth in the paper
\cite{greensite} (this remark concerns only abelian projected monopoles;
there exists another mechanism of confinement by $SU(N)/Z(N)$ dynamical
monopoles, see \cite{tombolius}). Anyway, we will not specify $Z(N)$
configurations in what follows, so that only their presence is important.

Going to display the phase structure of $SU(N)/Z(N)$ compact LGT,
one should remember that Lagrangian ? includes the adjoint characters
$\Omega (\partial p)$ \cite{mack2}. The fundamental Wilson loop
equals zero \cite{mack2}, implying the so-called "superconfinement"
of the static charges. There is a phase transition at a critical coupling
constant presumably of the first order, related to the condensation
of the $Z(N)$ monopoles \cite{hal} in this theory. The mixed theory
$S = \lambda_{1} \Omega^{adj} + \lambda_{2}\Omega^{fun}$ leads to the
picture of two phases existing at zero temperature:  the confining
phase with the area law behaviour, and the deconfining one.
Due to the absence of $Z(N)$ configurations in the bare
Lagrangian of $SU(N)/Z(N)$ LGT,
we cannot determine vortex potentials and consequently,
condensates as well, at least in the same manner. Certainly, they are
absent at the level of bare Lagrangian.  A similar situation takes
place in the positive plaquette model \cite{mack3} which eliminates
all thin $Z(N)$ vortices from the standard $SU(2)$ LGT. MC-simulations
indicate that string tension in this model is much less than in the
standard $SU(2)$. Since the Lie algebras of $SU(N), SU(N)/Z(N)$ and
positive plaquette model are the same, the naive continuum limits of
these models are the same as well, and equal the Yang-Mills action.
This means that $Z(N)$ configurations disappear from $SU(N)$ in this
limit.

Let us, therefore, display the role of the $Z(N)$ subgroup in the phase
structure of the compact $SU(N)$ model at a finite temperature.
We would also like to pay an attention to the chromoelectric part
of the model in this example. The partition function at a finite temperature
\begin{equation}
Z=\int D\mu (U_{n}) D\mu (U_{0}) \exp (\lambda \sum_{p} \Omega(\partial p)
+ \lambda_{0} \sum_{p_{0}} \Omega(\partial p_{0})),
\label{pfft}
\end{equation}
\noindent
where $p_{0}, (p)$ are time-like (space-like) plaquettes and
\begin{equation}
\lambda_{0} = \xi \frac{2N_{c}}{g^{2}}, \
\lambda = \xi^{-1} \frac{2N_{c}}{g^{2}}, \
\xi = \frac{a_{\sigma}}{a_t},
\label{coup}
\end{equation}
\noindent
is calculated at the following boundary conditions:
\begin{equation}
U_{\mu}(x,t) = U_{\mu}(x,t+N_{t}).
\label{pbc}
\end{equation}
\noindent
These conditions (\ref{pbc}) generate new physical degrees of freedom
which can be taken as the eigenvalues of the Polyakov loop \cite{pol2}
\begin{equation}
W_x = P\prod_{t=1}^{N_t}U_0(x,t).
\label{pl}
\end{equation}
\noindent
The compactness in the temporal direction leads to a $Z(N)$ global symmetry
of the model.
This means, multiplication of all links in the time direction
in the three dimensional $x,y,z$-torus by a $Z(N)$ element does not
change the action, though a single Polyakov loop transforms as
\begin{equation}
W_{x} \longrightarrow zW_{x},  \  z \in Z(N).
\label{trlaw}
\end{equation}
Thus, an expectation value of the Polyakov loop
can be used as an order parameter to measure a spontaneous breaking
of the $Z(N)$ symmetry. The corresponding phase transition
is well-known as the deconfining one \cite{sussk},
and in the high-temperature phase the $Z(N)$ symmetry was
spontaneously broken \cite{yaffe} (see, however, \cite{trl}).

What is the role of $Z(N)$ configurations at a finite temperature?
Let us sketch now some recognized results obtained
from the chromoelectric part of the action (\ref{pfft}), i.e. the term
$\lambda_{0} \sum_{p_{0}} \Omega(\partial p_{0})$ (to avoid possible
confusion we would like to stress that we are dealing with the Euclidean
formulation of LGT; ``chromoelectric part'' of the action means in this case
time-like plaquettes). Fixing the temporal gauge $\partial_{0} A_{0} = 0$
and performing the integration over the space gauge fields $U_{n}(x,t)$ we
come to the partition function of the form in the limit $a_t \rightarrow 0$
\begin{equation}
Z = \int \prod_{x}D\mu (W_x) \prod_{x,n}
{\sum_{l}\exp(-\gamma C_{2}(l))
\Omega_{l}(W(x))\Omega_{l}^{\ast}(W(x+n))},
\label{pfpl}
\end{equation}
\noindent
where $\gamma = g^{2}(2Ta_{\sigma})^{-1}$, $\Omega_{l}$
is the character of the l-th irreducible representation  of $SU(N)$
and $T = N_t a_t$ is the temperature. The partition
function (\ref{pfpl}) has been studied in numerous articles through
different approaches including MC-simulations.
There is a phase transition of the second order
for $SU(2)$ and of the first order for $SU(3)$ at some critical value
$N_{t} \lambda_{0}^{-1}$. The expectation value of the fundamental
character behaves according to
\begin{eqnarray}
\langle \{N^{-1}SpW_{x}\}\rangle = \left\{  \begin{array}{c}
	       0, \ T<T_{c}^{D},\ confinement\ phase,  \\
		  z \ast f(T),\ T>T_{c}^{D},\ z \in Z(N), \ f(T) \leq 1,
		   \\
	\ deconfinement\ phase. \end{array} \right.
\label{expv}
\end{eqnarray}
\noindent
Let us limit ourselves to $Z(N)$ subgroup in (\ref{pfpl}):
\begin{eqnarray}
\int D\mu (W) &\rightarrow & \frac{1}{N} \sum_{z \in Z(N)}, \nonumber  \\
\Omega_{l}(W) &\rightarrow & zd_{l}, \ if \  \Omega_{l} \rightarrow
z \Omega_{l},		\nonumber    \\
\Omega_{l}(W) &\rightarrow & d_{l}, \ if \  \Omega \ is\  \ invariant\
\ under\  Z(N),
\label{model}
\end{eqnarray}
\noindent
where $d_{l}$ is the dimension of $l$-th representation. The resulting
model has the same qualitative features as the initial one in (\ref{pfpl}).
Both deconfinement phase transition and confinement take place.
If we consider expansion for $W_{x}$ around the unit matrix (as it is made
at the naive continuum limit), we will lose this phase structure, because
the system stays in one of the minima of the $Z(N)$-broken deconfined phase.
Let us briefly summarize. $Z(N)$ configurations might play the crucial role
in the confinement mechanism.
It follows, if one wants to construct a noncompact theory which
could display this confinement mechanism it will be necessary
to include these configurations in such a theory. In fact, it is the old
problem. There were some attempts to implement $Z(N)$ configurations into
the Yang-Mills theory (see, for instance, \cite{stul} and references therein).
We would like to use an other method described in the following text.

\section{The essentials of noncompact LGT}

In order to show the principal difference between noncompact gauge
theories and compact ones, we shall briefly point out some
aspects of noncompact Yang-Mills theories on the lattice. They were
introduced first in \cite{patr} and studied intensively in
\cite{zwan,cah,pal,cah1} (see also references in \cite{cah1}).
The basic element is the gauge potential $A_\mu = A^{a}_{\mu} t^a$, where
$t^a$ are generators of the $SU(N)$ group.
The derivatives are represented by the finite-difference form
\begin{equation}
\partial_{\mu} f \rightarrow \frac{1}{a}[f(x+\mu) - f(x)],
\label{fd}
\end{equation}
\noindent
and integrals over the four-dimensional space are changed into sums
over all lattice sites. The path integral is defined as the integral over
all noncompact gauge fields $A_\mu (x)$ calculated in each lattice site.
The partition function is, therefore, defined by the relation:
\begin{equation}
Z = \int\ \prod_{x,\mu} dA_\mu (x) exp[-a^4 \sum_x \sum_{\mu,\nu}
(F_{\mu,\nu})^2].
\label{pffd}
\end{equation}
\noindent
The gauge invariance in these models is explicitly broken.
Hence, the gauge fixing mechanism represented by the Faddeev-Popov ansatz
cannot be directly simulated by the Monte-Carlo process. The method based
on simulation of a diffusion equation was proposed \cite{zwan}.
Gauge fixing is assured by introducing of the local gauge fixing force
to the diffusion equation, tangent to the gauge orbit. Expectation values
of gauge invariant quantities should be independent of such forces.
Despite the explicit breakdown of the gauge invariance,
in the limit $g^2 \rightarrow 0$ the asymptotic freedom is presented
\cite{patr,zwan}. One may suggest that the breakdown of the gauge invariance
on the level of the bare Lagrangian is not of crucial importance
due to restoration of the gauge symmetry in the expected region
$a \rightarrow 0$ of the quantum theory, since the terms that caused the
breakdown are proportional to the lattice spacing.
The main contribution to the path integral results from
a compact region defined by a gauge condition (local gauge force)
\cite{zwan}. No evidence for confinement was found. The string tension
vanished even at very strong values of the coupling constant.
The Wilson loop obeyed the perimeter law \cite{patr,zwan,cah}.
The expectation value of the Polyakov loop always differs from zero.
Similar behaviour was also observed in the theory with the gauge
$A_0=0$ \cite{patr}. From the point of view of these facts,
noncompact gauge theories resemble $SU(N)/Z(N)$ compact ones in the
weak coupling region, rather than $SU(N)$.

Some attempts to find a proper solution were connected to the fact
that the explicit violation of the gauge symmetry is the reason
for the absence of the confining forces \cite{pal,cah1}. However,
this opinion does not look to be well motivated. As we pointed out
earlier, in the quantum noncompact theory the asymptotic freedom
is observed and the contribution to the path integral results from the
compact region. Thus, there is, actually, no reason to believe
that the gauge invariance is not restored in the quantum theory after
taking the limit $a \rightarrow 0$. One more argument comes from the
consideration of the finite temperature behaviour of the noncompact model
(see section 4 of the present paper). If we believe in a common mechanism
of the confinement in the compact lattice gauge theory and in the noncompact
gauge theory then we should expect similar behaviour of
the Wilson (Polyakov) loops in the noncompact gauge model restricted to
the chromoelectric part as well. If we calculate the partition function
($SU(2)$ gauge group for simplicity)
\begin{equation}
Z = \int \prod_{x,\mu} dA_\mu(x,t) exp[-a^4 \sum_x (F_{0,\mu})^2]
\label{pfdt}
\end{equation}
\noindent
at the periodic boundary conditions $A_\mu(x,t) = A_\mu(x,t+N_t)$
in the temporal gauge, we shall find that $ \langle Tr W \rangle \neq 0$
at any temperature, perhaps with exception of the case of the
infinite coupling  $g^2 \rightarrow \infty$  \cite{zwan}.
Thus, the model (\ref{pfdt}) does not display confinement behaviour.
Let us emphasize that the final result is the gauge invariant as well as
the expression for the partition function after integration out of space
gauge fields $A_n(x)$ (see for technical details the next section).
This, together with independence of the MC results of the gauge and the
restoration of the asymptotic freedom, refutes the usual
objection concerning a connection between the vanishing of the string
tension and the breakdown of the gauge symmetry in noncompact models.

In fact, some kind of invariant integration over gauge fields is present
in all models constructed with the goal of avoiding this
explicit violation of the gauge invariance \cite{pal,cah1}. For instance,
the model proposed in \cite{pal} is equivalent to the dielectric LGT
introduced in \cite{mack4}.
Let us consider the $SU(2)$ Yang-Mills action with potentials \cite{pal}
\begin{equation}
A_{\mu}^{Y-M} = A_{\mu}^{a}t^{a} \rightarrow A_{\mu}^{d} =
A_{\mu}^{Y-M} + IT_{\mu},
\label{dpot}
\end{equation}
\noindent
where $T_{\mu}$ is a new noncompact potential proportional to a unit
matrix in colour space. Rewriting the obtained action
on the lattice in the finite-difference form we have as result the
dielectric theory \cite{mack4} since we can use the representation
$A_{\mu}^{d} = \rho_{\mu}(x)U_{\mu}(x)$  where $U_{\mu}(x) \in SU(2)$,
$0 \leq \rho < \infty$. We are allowed to choose the potential
for the dielectric field $\rho$ in such a form that \cite{pal}:
1) naive continuum limit equals the standard Yang-Mills action;
2) Wilson loops and corresponding string tension behave like those
in the compact Wilson model;
3) at the weak coupling asymptotic freedom exists. Thus, the theory is
noncompact but confinement of static charges takes place.
Since, however, the integration measure includes the invariant
measure of $SU(2)$ group, it can be the reason of the area law. Further,
introduce the following restriction for the noncompact field $T_{\mu}$
in (\ref{dpot}): $0 \leq T_{\mu} < \infty$.
It follows, that we should consider $SU(N)/Z(N)$ as the gauge
group since $U \in SU(N)/Z(N)$ in this case.
So, as we have discussed earlier
we have to use adjoint $SU(2)$ representation for gauge matrix $U$.
It means immediately that the fundamental Wilson loop equals zero
whereas the adjoint loop can show critical behaviour.

One more approach starting from a generating functional for noncompact
Yang-Mills theory has been discussed in Ref.\cite{cah1} where noncompact
fields are exposed to random compact gauge transformations (instead of the
Faddeev-Popov ansatz) at all lattice sites during every Monte-Carlo sweep.
Gauge invariance can be restored in this
approach and linearly rising potential has been observed. In the meantime
we cannot accept as conclusive, the original interpretation of the nature
of such behaviour. Indeed, random compact gauge transformations introduce
$Z(2)$ variables into the simulated theory, creating a reason for the
appearing of linear potential. We believe, in order to comprehend the
problem, the
supplementary MC-simulations with random gauge transformations $V(2)$
belonging to: 1) $Z(2)$ and 2) $SU(2)/Z(2)$ are highly desirable.
Then, what we would expect in the first case (no explicit gauge symmetry
restoration) is the confining behaviour in the strong coupling region
and the deconfining phase transition in the weak coupling region.
In fact, the basis for this is supported by the resemblance of the resulting
theory and the $Z(2)$ gauge theory. The Wilson loop behaviour in the full
range of coupling is less predictable for the second case. Surely, if we add
a summation over $Z(2)$ variables to the noncompact integration, the Wilson
loop will equal zero. We do not expect that the adjoint Wilson loop
will obey area law, at least in the region of weak coupling.

In our opinion, there are two different explanations of such behaviour
of noncompact lattice models:

1) The confinement mechanism in continuum gauge theory is different
from the one on the lattice (the so-called "light" confinement mechanism
which works only in the presence of the dynamical quarks \cite{polrev})
and this mechanism could work in noncompact lattice models.

2) Noncompact Yang-Mills theory belongs to the other universality class
without quark confinement.

The third possibility, namely that confinement is solely the property
of compact gauge theory is not confirmed by an example of the dielectric
gauge theory discussed above.
The following facts indicate that the second choice could be the right one:

i) There is no phase transition, depending on the coupling constant,
at zero temperature in pure $SU(2)$ and $SU(3)$ gauge
theories; it follows that a nonperturbative weak coupling limit in a bare
constant can define a noncompact model different from the naive lattice
Yang-Mills theory.

ii) The expectation value of the Polyakov loop differs from zero
in $SU(N)$ gauge theory if the vacuum is not invariant under
$Z(N)$ rotations\cite{pol}. The standard Yang-Mills theory with a flat
integration measure does not possess $Z(N)$ invariance.

iii) $SU(N)$ compact theory has $N$ global minima whereas noncompact
Yang-Mills theory has alone minimum. According to \cite{cah1},
such a periodicity could be responsible for
the confinement but all the minima with the exception of the trivial one
are nonphysical. We do not think that the argument of Ref. \cite{cah1} is
correct in this respect.

We believe, in full accordance with the results
\cite{patr,zwan,cah} discussed above, that the compact Wilson theory and
the noncompact lattice Yang-Mills theory belong to two different classes
of universality: we do not expect the Yang-Mills theory in its naive lattice
form to be able to describe confinement. Our method of discovering a proper
theory is to construct a noncompact model starting from the compact Wilson
model but not from the continuum Yang-Mills theory taking its naive lattice
form (\ref{pffd}). In order that the readers
have a guideline to the manuscript we present here a short description
of the main ideas. To secure a confining theory in the weak coupling region
we propose to execute the summation over $Z(N)$  variables in the compact
formulation and then to take a noncompact limit expanding the resulting
$SU(N)/Z(N)$ matrices around all minima of the effective action obtained.
Further, as is known from the studies of the strong coupling lattice
models, invariant group measure  (at least, for $A_{0}$ gauge field)
can be of great importance for the finite temperature confinement
(for the definition of the invariant measure, see section 5).
The flat integration measure for the $A_{0}$ field fails to respect
the $Z(N)$ global symmetry of the vacuum \cite{pol}.
Thus, the next step should be to
include the invariant measure contribution in the noncompact effective
model. Of course, the principal questions appearing here are the expansion
of $SU(N)/Z(N)$ gauge matrices in the points of the minima, and what form
can be used for the invariant measure.

\section{Noncompact model with compact $A_0$ integration}

As the first step, we are going to explore the $SU(2)$
noncompact model defined by Eq.(\ref{pfdt}). We  would
like to reexamine the continuum limit of the chromoelectric part of the
lattice action in order to include $Z(2)$ invariance and compact
$A_0$ integration in the finite-difference gauge theory.
We begin by rewriting the chromoelectric part
of the action (\ref{pfft}) using the following gauge transformations
of space gauge matrices $U_n(x)$ \cite{bor}
\begin{eqnarray}
U_n(x,t) \longrightarrow (V_x)^t U_n(x,t)(V_{x+n})^{-t},
\nonumber   \\
U_{n}^{+}(x,t) \longrightarrow (V_{x+n})^t U_{n}^{+}(x,t)(V_x)^{-t}, \ \
 n=1,...,d    \nonumber \\
\label{gt}
\end{eqnarray}
\noindent
where $V$ is the $U_0$ gauge matrix in the static gauge.
Next, the chromoelectric part becomes
\begin{eqnarray}
\lambda_0\ \sum_{p_0} \Omega(\partial p_0) \rightarrow \bar{S}(E)\ =
\frac{2a_n}{a_t g^2}\ \sum_{x,n}  \nonumber   \\
\left \{ \sum^{N_t-2}_{t=0} Sp[I-U_n(x,t)U^+_n (x,t+1)]
+\ Sp[I-U_n (x,N_t-1)W_xU^+_n (x,0)W^*_{x+n}] \right \}.
\label{cha}
\end{eqnarray}
\noindent
Computing the continuum limit of $\bar{S}(E)$
we consider that at the limit $a_t \rightarrow 0$ we have
$W_x \rightarrow \exp (i\beta\ g\ A_0(x))$ in the given gauge.
Assuming the smoothness of the $A_0$ field in the sense that
\begin{equation}
 W_xW^*_{x+n}\ \approx\ \exp (-i\beta g a_n\partial_n\ A_0(x))\ \approx\
 I\ -\ i\beta g a_n\partial_n\ A_0(x)\ +\ ...
\label{smooth}
\end{equation}
\noindent
by help of the definition
$\frac{\delta_{t,0}}{a_t}\ \rightarrow \ \delta(t)$ one obtains
\begin{equation}
\bar{S}_{con}(E)\ =\ \int d^3 x \int^\beta _0 dt \{
(\partial_t A_n)^2 + \delta(t)Sp[L_1 + L_2/a_t] \},
\label{chacon}
\end{equation}
\noindent
where
\begin{equation}
L_1\ = \ A_n(\partial_t A_n)\ +\ \beta(\partial_n A_0)\ (\partial_t A_n)
-\ A_n W(x) (\partial_t A_n) W^*(x),
\label{L1}
\end{equation}
\noindent
\begin{equation}
L_2\ =\ A^2_n\ +\ \frac{1}{2}\beta^2\ (\partial_n A_0)^2\ -\
A_n W(x) A_n W^*(x).
\label{L2}
\end{equation}
\noindent
The first term in Eq.(\ref{chacon}) corresponds to the gauge $A_0=0$.
However, this gauge is incompatible with periodic boundary conditions.
$A_0$ can be set equal to zero everywhere except at one singular
point \cite{kaj}. The second term in Eq.(\ref{chacon}) reflects this
fact. Using the decomposition
$$
W(x)\ \sim\  I\ +\ i\beta g A_0(x)\ +\ ...,
\nonumber
$$
\noindent
we can easily demonstrate that
the function at the $\delta$-symbol in Eq.(\ref{chacon}) corresponds to the
remaining terms in $(F_{0,n})^2$. Due to the preceding discussion we do
not use the last decomposition. We are going to study the partition function
$\bar{S}_{con}(E)$ in the finite difference formalism.
Because the gauge fixing in Eq.(\ref{chacon}) is
equivalent to the static one (i.e. $\partial_0A_0=0$), the results,
calculated in both gauges, are the same.
We have in Eq.(\ref{chacon}) the singular term $L_2/a_t$.
Hence, a regularization procedure is necessary to define it properly.
Applying the finite difference approximation in Eq.(\ref{chacon})
we discover through algebra the relation to the partition function
\begin{equation}
Z\ =\ \int \prod_x\ d\mu (W_x)\ \prod_{x,n}\ Z_{1,2}(W)\ Z_3(W)
\exp(-S^{(1)}),
\label{zf}
\end{equation}
\noindent
where
\begin{equation}
S^{(1)}\ =\ \frac{2a_n}{a_t}\beta^2\ \sum_x\ (\partial_nA_0(x))^2,
\ \  \partial_n A_0 (x)\ = \ A_0 (x+n)-A_0 (x).
\label{sf}
\end{equation}
$Z_{i}$ is the path integral over space gauge fields:
\begin{eqnarray}
Z_{1,2}\ &=& \ ( Det\ M^{bc}_{tt'} )^{-1/2}\ =  \nonumber  \\
& &\int\ \prod^{N_t-1}_{t=0}\ dA^1_n(t)\ dA^2_n(t)
\exp \{ -A^b_n (t)\ M^{bc}_{tt'}\ A^c_n (t')\ \},
\label{z0t}
\end{eqnarray}
\noindent
where $b,\ c\ =\ 1,2$ are the colour indices and
\begin{eqnarray}
Z_3\ = \ \int\ \prod^{N_t-1}_{t=0}\ dA^3_n (t)
\exp \{ -A^3_n (t)\ \tilde{M}_{tt'}\ A^3_n (t') \nonumber  \\
+\frac{2a_n^2}{a_t}\beta (\partial_n A_0 (x))  \sum^{N_t-1}_{t=0}
(\delta_{t,N_t-1}-\delta_{t,0})\ A^3_n (t)\ \} = \nonumber \\
\ ( Det\ \tilde{M}_{tt'} )^{-1/2}\ \exp \left [
\frac{2a^4_n}{a^2_t}\beta^2 (\partial_n A_0 (x))^2
[ (\tilde{M}_{0,0})^{-1}\ -\ (\tilde{M}_{0,N_t-1})^{-1} ] \right ].
\label{zt}
\end{eqnarray}
\noindent
We use the representation:
$$
M^{bc}_{tt^{\prime}}\ =\ \frac{a^3_n}{a_t}
\left( \begin{array}{lllll}
 2I    & -I   & \cdots	&   0	 &  m^{cb}     \\
-I     & 2I   & \cdots	&   0	 &   0	       \\
 0     &  0   & \cdots	&   0	 &   0	       \\
\vdots &\vdots & \vdots & \vdots & \vdots      \\
0      &  0    & \cdots &  2I	 & -I	       \\
m^{bc} &  0    & \cdots &  -I	 & 2I
\end{array} \right)
\label{det}
$$
and
$$
\tilde{M}\ =\ M^{bc}_{tt'}(m^{bc}=m^{cb}=I).
\label{mat}
$$
\noindent
$m^{bc}$ is the matrix $2\otimes 2$ constructed from the scalar
$Sp\ W(x)$ and the octet $Sp\sigma^3 W(x)$ parts of the Polyakov
loop. Through parametrization
\begin{equation}
W(x)=\exp (\frac{i\varphi(x)\sigma^3}{2}), \ \varphi(x)=\beta g A_0(x),
\label{param}
\end{equation}
\noindent
one finds
\begin{eqnarray}
m^{cb}=\left( \begin{array}{c}
		      \cos \varphi(x)\ \ -\sin \varphi(x) \\
		      \sin \varphi(x)\ \ \ \cos \varphi(x)
		\end{array} \right).
\label{comp}
\end{eqnarray}
\noindent
Introducing the notation
\begin{equation}
S_{eff}(A_0(x))=S^{(1)}-\frac{2a^4_n}{a^2_t}\beta^2\sum_{x,n}
(\partial_nA_0(x))^2 \left [ (\tilde{M}_{0,0})^{-1}\ -
\ (\tilde{M}_{0,N_t-1})^{-1} \right ],
\label{seff}
\end{equation}
\noindent
we can represent the partition function in the form
\begin{equation}
Z=[(Det\ \tilde{M}_{tt'})^{-1/2}]^{N^3_{\sigma}}\ \int e^{-S_{eff}(A_0(x))}
\prod_{x,n}(Det\ M^{cb}_{tt'})^{-1/2}\ \prod_{x} d\mu(W_x).
\label{zful}
\end{equation}
\noindent
It follows from (\ref{zt} - \ref{zful}) that the third color component of the
gauge field does not interact with the first and the second components
(because of chosen gauge).
However, only the third component leads to the second
term in (\ref{seff}). Since $\tilde{M}_{tt^{\prime}}$ does not depend on
$W(x)$ this contribution renormalizes a free theory $S^{(1)}$.
At $N_{t} \rightarrow \infty$ we have precisely
\begin{equation}
S_{eff} = \alpha \sum_{x,n}(\partial_{n}A_{0}(x))^{2}, \
\alpha = const \frac{a_{n}}{a_{t}} \beta^{2}.
\label{sful}
\end{equation}
\noindent
The determinant can be calculated to produce
\begin{equation}
DetM^{cb}_{t t^{\prime}} = (\frac{2a_{n}^{3}}{a_{t}})^{N_{t}}
\sin^{4}\frac{\varphi (x)}{2}.
\label{detf}
\end{equation}
\noindent
Since $d\mu(W_x) = \sin^{2} \frac{\varphi (x)}{2} d\varphi (x) $ we finally
obtain the partition function to be of the form (up to an irrelevant
constant)
\begin{equation}
Z \sim	\int e^{-S_{eff}(A_{0}(x))} \prod_{x}
\left [ \frac{\sin^{2} \frac{\varphi (x)}{2}d\varphi (x)}
{(\sin^{2} \frac{\varphi (x)}{2})^{d}} \right ],
\label{zncp1}
\end{equation}
\noindent
where $d$ is the space dimension.
It is clear from (\ref{zncp1}) that the singular contribution to (\ref{chacon})
is well-controlled in the present regularization (since there are no
time derivatives at the singular term and all singular terms are proportional
to $a_{t}^{-1}$). Obviously, the constructed theory will be equivalent
to (\ref{pfft}) (restricted to the chromoelectric part) when  the matrices
$U_{n}$ are expanded around unit matrices and $U_{0}$ are being kept
in their lattice form (it is also the proof that these calculations
do not depend on chosen gauge). To verify this we note that all
equations (\ref{zf}-\ref{zt}) are valid in this case and matrix
$M_{tt^{\prime}}^{cb}$ becomes
$$
M_{tt^{\prime}}^{cb} \rightarrow \overline{M}_{tt^{\prime}}^{cb} =
\left( \begin{array}{lllll}
 2I	& -m^{bc} & \cdots	&   0	 &  m^{cb}     \\
-m^{bc} & 2I	  & \cdots	&   0	 &   0	       \\
 0	&  0	  & \cdots	&   0	 &   0	       \\
\vdots	&\vdots   & \vdots & \vdots & \vdots	       \\
0	&  0	  & \cdots &  2I	 & -m^{bc}	\\
m^{bc}	&  0	  & \cdots &  -m^{bc}	 & 2I
\label{M}
\end{array} \right).
$$
\noindent
Since $DetM_{tt^{\prime}}^{cb} = Det \overline{M}_{tt^{\prime}}^{cb}$
we obtain the same result (\ref{zncp1}).

Let us now discuss this example comparing (\ref{zncp1}) with
(\ref{pfpl}). In parametrization (\ref{param}) the character expansion
in (\ref{pfpl}) can be expressed as
\begin{equation}
\prod_{x,n} \Omega_{l}(W_x)\Omega_{l}(W_{x+n}) = \prod_x
(\sin^{2} \frac{\varphi_{x}}{2})^{-d} \prod_{x,n}
\sin \frac{\varphi_{x}}{2}(2l+1)  \sin \frac{\varphi_{x+n}}{2}(2l+1).
\label{32}
\end{equation}
\noindent
We conclude from this that (\ref{pfpl}) and (\ref{zncp1})
are different theories. If we integrate over $d\mu (W_x)$ in (\ref{pfpl})
we will find that only closed loops contribute to the partition function
whereas there is no such property in (\ref{zncp1}). However, just this
property is the cause of linear potential at a low temperature.
Further, using the Poisson summation formula  applied to calculate
the sum over characters in (\ref{32}) we can easily show that at a high
temperature (lattice spacing $a$ is fixed)
\begin{equation}
\prod_{x,n} \sum_{l}\exp(-\gamma C_{2}(l))
\Omega_{l}(W_{x})\Omega_{l}(W_{x+n}) \longrightarrow_{T \rightarrow \infty}
\exp(-S_{eff} + S_{loc} + O(a))
\label{33}
\end{equation}
\noindent
with $S_{eff}$ to be of the form as in (\ref{sful}) \cite{sussk,su3},
$S_{loc}$ is a local function of $A_0$, and the measure in (\ref{zncp1})
should be treated as smooth function. The correlation function
of the Polyakov loops has a form corresponding to a screening potential
between probe quarks in this region. Hence, the model presented in
(\ref{zncp1}) is capable of describing only the high temperature
deconfined phase.
It follows immediately from this consideration that the pure Yang-Mills
theory (\ref{pfdt}) does not describe the confinement phase since we have
taken into account a larger number of gauge configurations (we did not expand
the Polyakov loops around unit matrix) generated by compact fields on
time-like plaquettes. We are convinced from this example that
(\ref{pfft}) and (\ref{pffd}) can belong to different universality
classes. We could obtain slightly more information if we did not presuppose
the smoothness of $A_{0}$ field in the sense (\ref{smooth}). The
smoothness of $A_{0}$ means that neighboring  ``spins'' $W_{x}$ and
$W_{x+n}$ are oriented approximately in one direction, while in the
confinement phase the configurations being essential for confinement
should be strongly  disordered (for example, $W_{x} \sim I$,
$W_{x+n} \sim -I$). In this case the expansion (\ref{smooth})
will be obviously invalid. If we had constructed the continuum limit in
the time direction only, we would obtain the effective action of the form
\begin{equation}
S_{eff} = \gamma \sum_{x,n} \cos(\varphi_{x} - \varphi_{x+n}).
\label{seff3}
\end{equation}
\noindent
This action coincides with the effective action for the $U(1)$ lattice
compact theory appearing in the Hamiltonian formulation at a finite
temperature. There is a phase transition from the low temperature
confining phase to the deconfining one in this theory.
However, this method of calculation is not mathematically well-founded
for the $SU(2)$ gauge group. If $W_{x} \sim I$
and $W_{x+n}$ is far from this configuration, then the expansion of
$U_{n}(x,t)$ around the unit matrix will not lead to the true minima
of the action. Two possible avenues to promote our calculations
are available in principle.
The first one consists of applying some conjectures proposed in
\cite{pol} and, then, in transition from theory (\ref{zncp1}) to
some effective model which should be calculated in the framework of
the renormalization group scheme.
There exists a concern that not all the important
configurations have been taken into account. The last example (\ref{seff3})
demonstrates that space gauge field configurations $Z \in U_{n}(x)$
could be essential for obtaining the true minima of the quantum theory.
In the next section we analyze the first of these possibilities.

\section{JLP - model and simulation of the invariant measure
contribution}

The next model we would like to examine was proposed
in \cite{pol} (see also \cite{polrev}) (in what follows we will call
the model: the JLP-model). The question is, how could one
simulate the contribution of the invariant group measure in a
noncompact (either lattice or continuum) theory.  A flat integration measure
fails to respect the $Z(N)$ global symmetry of the lattice action.
In this case the expectation value of the Polyakov loop differs from zero.
Usually, the definition of the invariant measure on $SU(N)$ group includes
a compact region of integration and weight function in the corresponding
integrand. In this section speaking about invariant measure we mean
only this weight function which is local contribution to the action of LGT.
The basic idea of the JLP-model
consists in the assumption that one should simulate this contribution
making use of a local $Z(N)$ invariant potential for $A_0$ gauge field.
Then, a symmetry argument suggests that the action of the confining $SU(2)$
noncompact model involves a non-polynomial periodic term depending
on the $A_{0}$ gauge field:
\begin{equation}
S_{Y.-M.} \rightarrow  \frac{1}{g^2} \int Sp (F_{\mu \nu})^2 d^4 x -
\frac{1}{a^4} \int d^4 x \ln ( \sin^2 A_0(x,t)ag),
\label{jpl1}
\end{equation}
\noindent
where, following \cite{pol}, we did not fix the static gauge but chose
the diagonal form for $A_{0}$. The basic assumption is that
the cutoff (lattice spacing in our case) of the theory is renormalizable
and is left finite in the continuum. If this is the case, the renormalized
Lagrangian is of the sine-Gordon form. Thus, we have for the action
\begin{equation}
S_{eff} = S_{Y.-M.} + \mu \sum_{m=1}^{\infty} \nu_m
\cos mTgA_0 = S_{Y.-M.} + V(A_0),
\label{jpl2}
\end{equation}
\noindent
where $\nu_m$ is an arbitrary coefficient in the model. The potential
$V(A_{0})$ is a sum over the characters of the $SU(2)$ gauge group
which are trivial on $Z(2)$.
The new constant $\mu$ can be interpreted as the so-called
hidden coupling constant \cite{topren} and should be calculated
in the course of the renormalization procedure. If one now takes another
assumption, namely that the dynamics of the space gauge fields is not
essential for the confinement, we will get the following effective theory
\begin{equation}
S_{eff} =  \int d^4 x (\frac{1}{g^2} Sp (\partial_{\mu}A_0(x))^2  -
V(A_0)).
\label{jpl3}
\end{equation}
\noindent
It has been claimed in \cite{pol} that the Wilson
loop obeys the area law and this leads to the linear
potential between probe quarks. The string tension appears to be
proportional to $\mu$. The model possesses global $Z(2)$ symmetry
which, however, appeared to be broken at any values of the coupling
constants $\mu$ and $\lambda$ \cite{gopfert}.
This seems to be in contradiction with
the main idea of \cite{pol}, since the invariant measure was introduced
to preserve the $Z(2)$ symmetry of the vacuum. In ref.\cite{gopfert}
it has been rigorously shown that the correlation functions of the kind
$<\sin A_0(0)/2 \sin A_0(R)/2>$ behave like those in the free scalar model.
This leads to the nonzero string tension $\alpha = \mu /\lambda$
if we choose the appropriate sign in $\mu$ (the effective action
(\ref{jpl3}) should have a maximum at $A_0=0$).

We studied a simplified version of the model (\ref{jpl3}) with
$\nu_m = \delta_{m,1}$. This approach is sufficient for our arguments
since the string tension in \cite{pol} is non-zero at this level.
Thus, we start from the partition function
\begin{equation}
Z = \int_{- \infty}^{\infty} \prod_x du_x e^{-S(u)},
\label{jpl4}
\end{equation}
\noindent
where the action is of the sine-Gordon type
\begin{equation}
S(u) = \lambda \sum_{x,x^{\prime}} u_x M_{x,x^{\prime}} u_{x^{\prime}}
- \mu \sum_x \cos u_x
\label{jpl5}
\end{equation}
\noindent
with
\begin{equation}
uMu = d u_x^2 - \sum_n u_x u_{x+n},
\label{jpl6}
\end{equation}
\noindent
and $d$ is the space dimension.
We are going to calculate the following correlation function
\begin{equation}
\Gamma (R) = < e^{i u(-R/2)/2} e^{-i u(R/2)/2} >,
\label{jpl7}
\end{equation}
\noindent
which can be interpreted as a correlation function of two Polyakov loops
in the static diagonal gauge (after we finished all calculations without
gauge fixing, we found that all results were essentially the same).
Introducing external sources $\eta_x = \frac{1}{2}(\delta_{x,-R/2} -
\delta_{x,R/2})$ into the partition function (\ref{jpl4}), we define
$\Gamma (R)$ as
\begin{equation}
\Gamma (R) = Z_{\eta}/Z,
\label{jpl8}
\end{equation}
\noindent
where
\begin{equation}
Z_{\eta} = \int_{- \infty}^{\infty} \prod_x du_x e^{-S(u) +
i\sum_x \eta_x u_x}.
\label{jpl9}
\end{equation}
\noindent
The corresponding potential between probe quarks is then
\begin{equation}
V(R) = - \ln \Gamma (R) - V_0,
\label{jpl10}
\end{equation}
\noindent
where $V_0$ is the self energy of two static charges. We want to investigate
two asymptotic regions on the plane $(\mu , \lambda)$: 1) $\mu \gg 1,
\lambda \gg 1$ and 2) $\mu \ll 1$. Let us begin with the first asymptotic.
Integrating the $u_x$ field, one obtains for $Z_{\eta}$ up to an irrelevant
constant
\begin{equation}
Z_{\eta} = (det \lambda M)^{-1/2} \sum_{l_x} \prod_xI_{l_x}(\mu)
\exp [-\frac{1}{\lambda} \sum_{x,x^{\prime}}
(l_x + \eta_x) M_{x,x^{\prime}}^{-1} (l_{x^{\prime}} + \eta_{x^{\prime}})].
\label{jpl11}
\end{equation}
\noindent
Here, $I_l$ is the modified Bessel function. Taking its asymptotic behaviour
at $\mu \gg 1$ we can make use of the Poisson summation formula to calculate
the sum over $l_x$ in (\ref{jpl11}). After this procedure one arrives
at the equation for the potential, to be of the form in the limit
$N \rightarrow \infty$, where $N$ is the number of lattice sites
\begin{equation}
V (R) = q(R) + \frac{1}{\lambda} \int d^3 k \sin^{2}\frac{k_{\sigma}
R_{\sigma}}{2} (\frac{1}{M_k} - \frac{1}{M_k + \frac{2\mu}{\lambda}}) -
V_0,
\label{jpl12}
\end{equation}
\noindent
where we have denoted
\begin{equation}
q(R) = \frac{1}{\lambda} \sum_{x,x^{\prime}}
\eta_x M_{x,x^{\prime}}^{-1} \eta_{x^{\prime}}
\label{jpl13}
\end{equation}
\noindent
and
\begin{equation}
M_k =  d - \sum_{\sigma =1}^d \cos k_{\sigma}.
\label{jpl14}
\end{equation}
\noindent
Calculating the right-hand side of eq.(\ref{jpl12}) we find the potential
of the general form for $R \rightarrow \infty$
\begin{equation}
V(R) \sim \frac{a}{R} - \frac{be^{-mR}}{R}
\label{jpl15}
\end{equation}
\noindent
with $m = \frac{2\mu}{\lambda}$ and $a$, $b$ are $R$-independent constants.

Considering asymptotic $\mu \ll 1$ (just this case corresponds to
the regime of \cite{pol}), it is convenient to rewrite $Z_{\eta}$
in an equivalent form as
\begin{equation}
Z_{\eta} = e^{-q(R)} \int_{- \infty}^{\infty} \prod_x du_x
\exp[- \lambda \sum_{x,x^{\prime}} u_x M_{x,x^{\prime}} u_{x^{\prime}}
- \mu \sum_x \cos (u_x + iQ_x)],
\label{jpl16}
\end{equation}
\noindent
where $Q_x = \frac{2}{\lambda} \sum_{x^{\prime}}
M_{x,x^{\prime}}^{-1} \eta_{x^{\prime}}$. In the first order in $\mu$
we find the potential $V(R)$ to be
(taking into account contribution of $V_0$):
\begin{equation}
V(R) = -\frac{1}{2}M^{-1}_{R/2,-R/2} + \mu \sum_x W_x(R).
\label{jpl17}
\end{equation}
\noindent
We introduced here
\begin{equation}
W_x(R) = 4 e^{-\frac{1}{\lambda}M^{-1}_{x,0}} \sinh
\frac{M^{-1}_{x,-R/2}}{2\lambda} \sinh \frac{M^{-1}_{x,R/2}}{2\lambda}
\cosh \frac{M^{-1}_{x,R/2} -M^{-1}_{x,-R/2} }{2\lambda}.
\label{jpl18}
\end{equation}
\noindent
If we consider asymptotic $\lambda \gg 1$ we can approximately represent
\begin{equation}
W_x(R) \approx \frac{M^{-1}_{x,-R/2}M^{-1}_{x,R/2}}{\lambda^2}.
\label{jpl19}
\end{equation}
\noindent
Eq.(\ref{jpl18}) with asymptotic behaviour (\ref{jpl19}) corresponds,
in our lattice notations, one-to-one to the result of \cite{pol} where the
confining potential has emerged from the term $\sum_xW_x(R)$. We performed
both analytical and numerical evaluations of the sum (\ref{jpl17}) in
the approach
\begin{equation}
W_x(R) \approx e^{-\frac{1}{\lambda}M^{-1}_{x,0}}
\left[ \frac{M^{-1}_{x,-R/2}M^{-1}_{x,R/2}}{\lambda^2} \right].
\label{jpl20}
\end{equation}
\noindent
Actually this sum is divergent, but it is decreasing as a function of $R$
for any finite number of lattice sites. We did the computations for various
values of $\lambda \geq 1$ and for $N = 20^3, 30^3, 40^3$. Certainly,
if we took at this stage the continuum limit in (\ref{jpl19}) and followed
the procedure of \cite{pol} we would get the linear potential. But we do not
think that this procedure is well founded. We considered the limit $N=\infty$
and introduced a different regularization to compute $V(R)$.
$\sum_xW_x(R)$ can be represented in the approach (\ref{jpl19}) as
\begin{equation}
W_x(R) \approx	 lim_{\epsilon \rightarrow 0}
(-\frac{\partial}{\partial \epsilon} \sum_{k_{\sigma}}
\frac{e^{i k_{\sigma}R_{\sigma}}}{d + \epsilon - \sum_{\sigma=1}^d
\cos k_{\sigma}}).
\label{jpl21}
\end{equation}
\noindent
The potential can be found after subtraction of the $R$-independent
divergences from the last equation. Indeed, increasing with $R$ potential
appears to have a negative sign. To achieve a confining potential we
have to choose $\mu < 0$. In \cite{pol} $\mu$ enters the effective action
just with this sign. We were not able to prove that this result is
independent of the regularization scheme for evaluations of these divergent
sums. Besides, there is an additional term
$e^{-\frac{1}{\lambda}M^{-1}_{x,0}}$ in $W_x(R)$ which was missed in
\cite{pol}. This term improves the convergence of the whole sum
but the renormalization procedure becomes even more complicated.

A more reliable way to calculate (\ref{jpl16}), in our opinion, is to use
the saddle point method. This method leads,
in the asymptotic under consideration, to the potential which has
the form (\ref{jpl15}) (we omitted all calculations since they are quite
transparent). This result is in agreement with rigorous results of
\cite{gopfert}. If the correlation function $<\sin A_0(0)/2 \sin A_0(R)/2>$
is exponentially decreasing and $Z(2)$ global symmetry is broken
then the correlation function $<\cos A_0(0)/2 \cos A_0(R)/2>$
is close to unity. Consequently, the correlation function (\ref{jpl7})
is close to unity as well, which implies a nonconfining potential.

We would like to summarize the main consequences and to provide some comments
on the reliability of this approach to the confinement in noncompact models.
Taking suitable correlation functions (see theorem 4 in \cite{gopfert}),
the nonzero string tension can be found even in the continuum limit
after proper renormalization procedure. If we may interpret
these correlation functions as those of the Polyakov loops in the effective
model, then we have confinement of static charges. Nevertheless, $Z(2)$
global symmetry is broken at all couplings. This type of confinement
resembles that of the $U(1)$ Villain lattice model investigated in
\cite{gopfert}. The sine-Gordon model is there an effective model of
the lattice abelian theory with Villain action. Thus, the first question is
whether this mechanism can reproduce the specific features of confinement
of the $SU(2)$ Wilson model. In this model $Z(2)$ global symmetry is unbroken
at zero and at low temperatures. A broken $Z(2)$ implies a screening
potential between probe quarks, and, thus a deconfinement phase. Thereby,
we have to use (\ref{jpl7}) as correlations of the Polyakov loops
in this model since, in any other case, it is unclear how to deal with the
deconfinement transition if we have the linear potential in the $Z(2)$
broken phase. In our opinion, to reproduce the specific features of the
$SU(2)$ Wilson theory we have to preserve not only the global center
symmetry, but also the local center symmetry.
In principle, the global symmetry
can be spontaneously broken as it happens in the standard sine-Gordon model,
whereas there should be no such breakdown at zero temperature $SU(N)$
models. To achieve the above stated goal,
it is not sufficient to introduce the invariant measure
into effective action. We present a modification of the JLP model which
respects {\it local} $Z(N)$ symmetry in \cite{ncmpt2}.

The second question concerns the assumption that the dynamics of space gauge
potentials is not essential for the confinement. This may not be the case,
and we have discussed in the previous chapter that $Z(2)$ configurations
contained in the compact lattice field $U_n(x)$ can be of great
importance. Let us consider the compact formulation at zero temperature.
If we set $U_n(x) = I$ everywhere, we get as a result an $XY$ model for
the $U_0(x)$ gauge field with an integration over the compact $SU(2)$ measure.
There is a phase transition in this model implying deconfinement of
static quarks. However, no phase transition should take place
in $SU(2)$ at zero temperature. Hence, we are not allowed to neglect
dynamics of space gauge potential, at least in this naive form.
Further, it is obvious from (\ref{zncp1}) that the noncompact
integration over space potentials can significantly change the effective
integration measure for the $A_0$ gauge field because the determinant
(\ref{detf}) generates a local contribution to the measure. Moreover, we can
see from (\ref{zful}), (\ref{detf}) that the noncompact integration over
space gauge potentials generates just the invariant measure. On the other
hand, the contribution of the sine-Gordon type can appear not only from
the invariant measure but also from the effective action (\ref{33}).
The calculation of $S_{loc}$ in (\ref{33}) shows that this term is proportional
to the $\cos agA_0$ up to the corrections $O(a)$. The situation
becomes even more complicated when we consider the chromomagnetic part
of the action. The invariant measure can be cancelled completely in this
case as has been shown in \cite{boh} (in fact, there is currently no
common opinion on the cancellation of the invariant measure - see for
discussion \cite{seiler}). Hopefully there should be no such cancellation
at zero temperature. Here, the invariant measure can be included into
the effective action together with a compact measure for $Z(N)$ space gauge
configurations.

\section{Compactness and noncompactness in confinement:
discussion of simple models}

In this section we discuss the problem of compactness
and its importance regarding confinement.
In the broad class of lattice models, the compactness of the potentials
entering the original action, is an essential condition for
confinement. On the other hand, the compactness itself does not lead to
the linear potential. Indeed, the compactification performed following
the scheme
$$
\int_{-\infty}^{\infty} d\phi f(\phi) = \int_{0}^{2 \pi} d\phi F(\phi), \
F(\phi) = \sum_{j=-\infty}^{\infty} f(2 \pi j + \phi)
$$
\noindent
can change nothing in the correlation functions, so we adduce the following
example when the transition to the compact theory provides confinement.
We start from the theory of a scalar noncompact field in continuum with
the action
\begin{equation}
S = J_0 \frac{1}{2}\int d^4 x \sum_n (\partial_n \phi(x))^2 +
J_1 \int d^4 x \cos \phi(x),
\label{ex1}
\end{equation}
\noindent
where $n = 1,...,d$ and represent it on the lattice as
\begin{equation}
S^{lat} = J_0^{l} \frac{1}{2} \sum_{x,n} (\Delta_n \phi(x))^2 +
J_1^{l} \sum_x \cos \phi(x),
\label{ex2}
\end{equation}
\noindent
where $J_0^{l}$ and $J_1^{l}$ is connected with $J_0$ and $J_1$.
The compactified version of (\ref{ex2}) has the form
(up to an irrelevant constant and up to $a^2$ in lattice spacing $a$)
\begin{equation}
S^{lat}_{compt} = \sum_{x,n} (J_{-} \cos \frac{\phi_{x+n} - \phi_x}{2} +
J_{+} \cos \frac{\phi_{x+n} + \phi_x}{2}).
\label{ex3}
\end{equation}
\noindent
This expression coincides with the effective three-dimensional action
for $SU(2)$ gluodynamics at a finite temperature in the strong coupling
limit provided that $J_{-} - J_{+} \sim 0$ (see, for instance, \cite{su2gl}).
One can find linear potential in this model at small $J_{-}$ ($d = 3$)
and deconfinement transition to a phase with screening potential when
$J_{-}$ is increasing. What lesson may we extract from this?
The model (\ref{ex3}) is a version of the well-known
three-dimensional $XY$ model which displays a phase transition from
strong to weak coupling behaviour. At the small value of $J_{-}$
(low temperature strong coupling region of finite temperature $SU(2)$
theory or high temperature region of the spin system) correlation functions
fall exponentially and system is in the disordered phase. Regarding $SU(2)$
language, it means the area law for the Wilson loops. This behaviour
is due to the vortex loops which percolate through the lattice
\cite{xyvortex} (or become fat because of condensation, in other
terminology). In the weak coupling phase only short noninteracting
vortex loops are allowed, thus the behaviour of the system is mainly
defined by the contribution of the spin waves. Neglecting entirely
the vortex contribution in this region, we come to the free theory
of the scalar field on the lattice. The theories
(\ref{ex1}),(\ref{ex2}) are in the same universality class
while the theory (\ref{ex3})  belongs to an other class, although
it originates from (\ref{ex1}) and has only this naive
continuum limit. Thereby, the question is whether it is possible that
the theory (\ref{ex3}) could define a new continuum theory
with a disordered phase caused by vortex condensation? In other words,
can we construct a continuum limit of the model (\ref{ex3})
at the small value of $J_{-}$ (strong coupling region)? We investigate
this problem in the next paper \cite{ncmpt2}.

Our next example is related to the fact that a form of the effective
variables in the action can also be essential for confinement.
To illustrate this suggestion by way of a solvable model we need to restrict
ourselves to models which can be effectively reduced (for
dominating configurations) to the Gaussian integrals after
changing variables $\phi_x \rightarrow u(\phi_x) = u_x$ and expanding
the action around the main minimum. Thus, we consider the action
\begin{equation}
S_{eff} = - \frac{\beta}{2} \sum_{x,x^{\prime}} u_x
M_{x-x^{\prime}} u_{x^{\prime}} + \sum_x V(u_x) + i \sum_x
\eta_x \phi_x(u_x),
\label{ex4}
\end{equation}
\noindent
supposing that the potential $V(u_x)$ includes the contributions
both of the invariant measure and of the Jacobian of the substitution
$\phi_x \rightarrow  u_x$. The integration in the partition function
is performed with the measure $\prod_x du_x$ over the entire noncompact
region. When $u_x=\phi_x$ we have a model which is close to the JLP model
if $V$ represents the periodic potential. Let us suppose that a proper
effective variable is $u_x=e^{i\phi_x}$. Then we immediately obtain
\begin{equation}
e^{-F(R)} = <u_0u_R^{\ast}> \approx \frac{e^{-mR}}{R},
\label{ex5}
\end{equation}
\noindent
where
\begin{equation}
m = \frac{1}{N} \sum_x(\frac{\beta}{2}M_x + \frac{V^{\prime \prime}(0)}{2})
\label{ex6}
\end{equation}
\noindent
is the string tension in this model. Formally, the principal point
in this calculation was the transition from $\Gamma (R)=<\phi_0 \phi_R>$
to $\Gamma (R)=<u_0u_R^{\ast}>$ when $u_x=e^{i\phi_x}$. Let
us imagine that the original variables $\phi_x$ were compact
variables. Then, this rather trivial example demonstrates that in some
models with compact variables it is sufficient to perform only
noncompact Gaussian integration over the dominating configurations to
achieve the linear potential.

Our next example concerns the noncompact model presented in
\cite{stul,stul1}. This model is of the same spirit like \cite{mack},
where a sufficient condition for the
confinement was derived (see our discussion in the section 2).
It has been proven that if probability distribution of the vortices in the
compact
model obeys the area law it will lead to the disordering behaviour of
the Wilson loop and, consequently, to the confining potential. This
statement can be generalized for the noncompact model calculated in
\cite{stul} (see \cite{vortcond}). Unfortunately, both the theory
\cite{stul} and this generalization of the $Z(N)$ vortex confinement
mechanism to the noncompact models \cite{vortcond} appear very formal.
As we pointed out in section 2, the theory of \cite{stul} is noncompact and
not a naive limit of the Wilson model. It includes, in addition to Yang-Mills
potentials, singular $Z(N)$ transformations performed over two-dimensional
closed surfaces. The corresponding path integral contains a summation over
all possible two dimensional surfaces and determinants in the external
singular fields. It is unclear at the moment whether it is possible
to execute all these summations and to calculate the corresponding
determinants (it is the main reason for not adducing
any calculations here with discussed theory).
As such, we think this model demonstrates that $Z(N)$ variables
can be included in the noncompact limit of the compact theory.

The last example concerns a mathematical origin of confinement in the Wilson
theory and how its origin can be reproduced in noncompact models \cite{lt}.
We start from the finite temperature partition function for $SU(N)$
gluodynamics, obtained within the approach of time-like plaquettes, as
\begin{equation}
Z = \int \prod_{x}D\mu_x \prod_{x,n} \sum_{l} K_l(\gamma)
\Omega_{l}(x)\Omega_{l}^{\ast}(x+n).
\label{ex7}
\end{equation}
\noindent
Here, $D\mu_x$ is the invariant integratiom measure and $\Omega_{l}(x)$
is the character of the $l$-th irreducible representation.
Performing the invariant integration in the partition function we can easily
check that closed loops only contribute to the partition function.
For the correlation function $<\Omega_f(0)\Omega_{f}^{\ast}(R)>$
($f$ marks fundamental representation) we find out that the first nontrivial
term, surviving the invariant integration in the region
$K_l(\gamma) \rightarrow 0$ (low temperature), is the shortest path between
points $0$ and $R$ on the lattice:
\begin{equation}
<\Omega_f(0)\Omega_{f}^{\ast}(R)> \approx (K_f(\gamma))^R + O(K(\gamma)).
\label{ex8}
\end{equation}
\noindent
Could this picture be reproduced for noncompact fields? The positive answer
becomes straightforward supposing that we have the following partition
function for the noncompact field $u_x$
\begin{equation}
Z = \int_{-\infty}^{\infty} \prod_{x}[du_x e^{V(u_x)}] e^{\tilde{S}(u)}
\label{ex9}
\end{equation}
\noindent
and $e^{\tilde{S}(u)}$ can be expanded as
\begin{equation}
e^{\tilde{S}(u)} = \prod_{x,n} \sum_{l} C_l(\gamma) L_{l}(u_x)L_{l}(u_{x+n}),
\label{ex10}
\end{equation}
\noindent
where functions $L_l$ form the complete orthonormal basis
in the space of quadratically integrable functions with the weight
$e^{V(u_x)}$
\begin{equation}
\int_{-\infty}^{\infty} du e^{V(u)} L_l(u)L_k(u) = \delta_{l,k}.
\label{ex11}
\end{equation}
\noindent
Mathematically it indicates the same property as above: closed loops only
contribute to the partition function (\ref{ex9}). Calculating the correlation
function $<L_{k}(0)L_{k}(R)>$ we find the linear potential in a similar
manner as in the model with compact invariant integration:
\begin{equation}
<L_{k}(0)L_{k}(R)> \approx (C_k(\gamma))^R + O(C(\gamma)),
\label{ex12}
\end{equation}
\noindent
if the following equation is fulfilled
\begin{equation}
\int_{-\infty}^{\infty} du e^{V(u)} L_l(u) = 0.
\label{ex13}
\end{equation}
\noindent
Two points should be stressed here.
The weight $e^{V(u)}$ plays a role
of the invariant measure of the compact model. It enters the action
$S=V(u)+\tilde{S}$ as the local potential. The role of the $Z(N)$ symmetry
is to pick up those functions $L_k$ within the complete basis $\{ L \}$
which satisfy eq.(\ref{ex13}).
It is clear from the procedure above that this method could be directly
applied to the full Wilson action with plaquette interaction. We should
take a product of the $L_l$ functions in this case along the perimeter of
the minimal plaquette with the same weight, and to sum over $l$ in
(\ref{ex10}) obeying (\ref{ex13}).

Finally, the last question has to be answered: whether
the partition function (\ref{ex9}) may correspond
to any quantum field theory with acceptable properties besides
confining ones? In the paper  \cite{ncmpt2}
we try to synthesize all the essential results of this
paper into a general picture and present an investigation of
the $SU(2)$ compact Wilson model in the region of weak coupling.
We shall demonstrate that models of such types (\ref{ex9})
can be defined as a noncompact limit of the Wilson theory
in the weak coupling region if we execute summation over $Z(N)$ variables.
We shall calculate an effective noncompact model and prove
its confining behaviour.

We would like to thank M.~Hasenbush who has drawn our attention to the
fact that $Z(2)$ global symmetry is broken in the sine-Gordon model
at all couplings in $d>2$ dimensions. We are grateful to J.~Polonyi
and K.~Seiler for discussion of many points concerning the invariant
measure in the compact LGT and in the continuum theory.
The valuable remarks of N.~Kawamoto and D.~Miller on different stages
of this work are appreciated. Our special thanks
are for D.~Southworth and \v S.~Olejnik for the help in the preparing
of this manuscript.

The work was partially supported by NATO Linkage Grant No.930224.


\begin{thebibliography}{99}

\bibitem{wilson} K.~Wilson, Phys.Rev. D10 (1974) 2445.
\bibitem{pol} K.~Johnson, L.~Lelouch, J.~Polonyi, Nucl.Phys. B367
(1991) 675; J.~Polonyi,  \\
The confinement and localization of quarks, Proceedings
''Effective field theories of the standard model'', edited by Ulf-G
Mei\ss ner, Word Scientific 1991, Singapore, pp. 337--357
\bibitem{polrev} J.~Polonyi, F.~Csikor, A.~Patk\` os, K.~Szlachanyi,
''Selected topics in quark confinement'', Lectures, E\" otv\" os University,
Budapest, September 1992.
\bibitem{seil} K.~Osterwalder, E.~Seiler, Ann. Phys. 110 (1978) 440.
\bibitem{mack} G.~Mack, V.~Petkova, Ann.Phys. 125 (1980) 117.
\bibitem{munst} G.~M\" unster, Nucl.Phys. B180 (1981) 23.
\bibitem{mw}  R.D.~Mawhinney, Nucl.Phys. B321 (1989) 653.
\bibitem{mack2} G.~Mack, Phys.Lett. B78 (1978) 263.
\bibitem{stul} R.L.~Stuller, preprint BNL-41677 (1988).
\bibitem{stul1} R.L.~Stuller, preprint BNL-41678 (1988).
\bibitem{greensite}  L.~Del Debbio, M.~Faber, J.~Greensite,
Large-$N$ and strong coupling vs. dual superconductivity,
Proceedings 'Lattice-93',
in Nucl.Phys. B (Proc. Suppl.) 34 (1994) 237;
Nucl.Phys. B414 (1994) 594.
\bibitem{tombolius} E.T.~Tomboulis, Dynamical Monopoles and Confinement,
Proceedings Quarks'92, Edit by D.Yu.~Grigoriev, V.A.~Matveev,
V.A.~Rubakov, P.G.~Tinyakov, World Scientific,
Singapore-New Jersey-London-Hong Kong, p. 95-108;
$SO(3)$ monopoles as confinement mechanism, Proceedings 'Lattice-93',
in Nucl.Phys. B (Proc. Suppl.) 34 (1994) 192.
\bibitem{ncmpt2} O.A.~Borisenko, V.K.~Petrov, G.M.~Zinovjev and
J.~Boh\' a\v cik,
Phase structure and confinement properties of noncompact gauge theories.
II. Noncompact limit of $SU(2)$ compact gauge theory.
\bibitem{hal} I.G.~Halliday, A.~Schwimmer, Phys.Lett. 101B (1981) 327;
102B (1981) 337 (and references therein).
\bibitem{mack3} G.~Mack, E.~Pietarinen, Nucl. Phys. B205 (1982) 142.
\bibitem{pol2} J.~Polonyi, H.W.~Wyld, University of Illinois preprint,
ILL-85-83, 1985; \\
J.~Polonyi, in Advances in Lattice Gauge Theory, ed. by O.~Duka, J.~Owens,
1985.
\bibitem{sussk} A.M.Polyakov, Phys.Lett. 72B (1978) 477;
L.Susskind, Phys.Rev. D20 (1979) 2610.
\bibitem{yaffe} B.~Svetitsky, L.G.~Yaffe, Nucl. Phys. B210 (1982) 423;
 L.McLerran, B.Svetitsky, Phys.Rev. D24 (1981) 450;
J.~Kuti, J.~Polonyi, K.~Szlachanyi, Phys.Lett. 98B (1981) 199.
\bibitem{trl} M.Faber, O.A.Borisenko, G.M.Zinovjev,
Triality in QCD at zero and finite temperature: a new direction,
to appear in Nucl.Phys. B  (1995); hep-ph-9504264 (1995).
\bibitem{patr} A.~Patrascioiu, E.~Seiler, I.O.~Stamatescu,
Phys.Lett. B107 (1981) 364.
\bibitem{zwan} E.~Seiler, I.O.~Stamatescu, D.~Zwanziger,
Nucl.Phys. B239 (1984) 177.
\bibitem{cah} K.~Cahill, S.~Prasad, R.~Reeder, B.~Richter,
Phys.Lett. B181 (1986) 333.
\bibitem{pal} F.~Palumbo, M.I.~Polikarpov, A.I.~Veselov,
Phys.Lett. B297 (1992) 171; F.~Palumbo, Phys.Lett. B244 (1990) 55.
\bibitem{cah1} K.~Cahill, Gauge invariance and confinement in
noncompact simulations of $SU(2)$, Proceedings 'Lattice-93',
in Nucl.Phys. B (Proc. Suppl.) 34 (1994) 231;
K.~Cahill, G.~Herling, Noncompact gauge-invariant simulations
of $U(1)$, $SU(2)$, and $SU(3)$, Proceedings 'Lattice-94',
in Nucl.Phys. B (Proc. Suppl.) 42 (1995) 858.
\bibitem{mack4} G.~Mack, Nucl.Phys. B235 (1984) 197.
\bibitem{bor} O.A.~Borisenko, V.K.~Petrov, Yad.Phys. 54 (1991) 1708.
\bibitem{kaj} K.~Kajantie, Hot gluon matter and physical gauges,
Proceedings 'Physical and Nonstandard Gauges', ed. by P.~Gaigg,
W.~Kummer, M.~Schweda, Springer, Berlin (1990) 263.
\bibitem{su3} O.A.Borisenko, V.K.Petrov, G.M.Zinovjev,
Yad.Fiz. 45 (1987) 1115.
\bibitem{topren} J.~Polonyi, Proceedings 'Lattice-90',
in Nucl.Phys. B (Proc. Suppl.) 20 (1991) 32.
\bibitem{gopfert} M.~G\"{o}pfert, G.~Mack, Com.Math.Phys. 81 (1981) 97;
82 (1982) 545.
\bibitem{boh} J.~Boh\'a\v cik, Phys.Rev. D42 (1990) 3554.
\bibitem{seiler} K.~Seiler, A.~Sch\'{a}fer, W.~Greiner,
Effect of the Haar measure on the finite temperature effective potential
of $SU(2)$  Yang-Mills theory, preprint hep-ph-9502234.
\bibitem{su2gl} O.A.~Borisenko, V.K.~Petrov, G.M.~Zinovjev,
Theor.Mat.Fiz. 80 (1989) 381.
\bibitem{xyvortex} A.~Hulsebos, The behavior of vortex loops in the
$3-d$ $XY$ model, Proceedings 'Lattice-93',
in Nucl.Phys. B (Proc. Suppl.) 34 (1994) 695.
\bibitem{vortcond} O.A.~Borisenko, Confinement by means of $Z(N)$
vortex condensation in noncompact theories, in preparation.
\bibitem{lt} O.A.~Borisenko, V.K.~Petrov, G.M.~Zinovjev,
Proceedings 'Lattice-94', in Nucl.Phys. B (Proc. Suppl.) 42 (1995) 466.

\end{thebibliography}
\end{document}